# Nanoscale

## ARTICLE

# The Effects of Exfoliation, Organic Solvents and Anodic Activation on Catalytic Hydrogen Evolution Reaction of Tungsten Disulfide

Wanglian Liu,[a, b] John Benson,[c] Craig Dawson,[c] Andrew Strudwick,[c] Arun Prakash Aranga Raju,[c] Yisong Han, [a] Meixian Li,*, [b] and Pagona Papakonstantinou*, [a]

The rational design of transition metal dichalcogenide electrocatalysts for efficiently catalyzing hydrogen evolution reaction (HER) is believed to lead to the generation of a renewable energy carrier. To this end our work has made three main contributions. At first, we have demonstrated that exfoliation via ionic liquid assisted grinding combined with gradient centrifugation is an efficient method to exfoliate bulk $WS_2$ to nanosheets with a thickness of a few atomic layers and lateral size dimensions in the range of 100 nm to 2 nm. These $WS_2$ nanosheets decorated with scattered nanodots exhibited highly enhanced catalytic performance for HER with an onset potential of -130 mV vs. RHE, an overpotential of 337 mV at 10 mA cm$^{-2}$ and a Tafel slope of 80 mV dec$^{-1}$ in 0.5 M $H_2SO_4$. Secondly, we found a strong aging effect on the electrocatalytic performance of $WS_2$ stored in high boiling point organic solvents such as dimethylformamide (DMF). Importantly, the HER ability could be recovered by removing the organic (DMF) residues, which obstructed the electron transport, with acetone. Thirdly, we established that the HER performance of $WS_2$ nanosheets/nanodots could be significantly enhanced, by activating the electrode surface at a positive voltage for a very short time (60 s), decreasing the kinetic overpotential by more than 80 mV at 10 mA cm$^{-2}$. The performance enhancement was found to arise primarily from the ability of a formed proton-intercalated amorphous tungsten trioxide (a-$WO_3$) to provide additional active sites and favourably modify the immediate chemical environment of the $WS_2$ catalyst, rendering it more favorable for local proton delivery and/or transport to the active edge site of $WS_2$. Our results provide new insights into the effects of organic solvents and electrochemical activation on the catalytic performance of two-dimensional $WS_2$ for HER.

## Introduction

Due to growing concerns about meeting global energy demands whilst minimizing environmental pollution, hydrogen ($H_2$) is currently being actively investigated as a clean and renewable energy carrier for the replacement of fossil fuels.[1,2] Hydrogen is also a chemical commodity for some major industrial processes, including synthetic fertilizer or hydrocarbon productions and is also a useful product for driving key selective transformations, including carbon dioxide ($CO_2$) reduction for environmental remediation.[3] Today, hydrogen production has been dominated by steam reforming of natural gas such as methane, a high temperature process, which involves release of 1 molecule of $CO_2$ for every four molecules of $H_2$.[4] A clean and sustainable synthesis of hydrogen can be achieved through water electrolysis ($2H_2O \rightarrow 2H_2 + O_2$), when combined with renewable power sources such as solar or wind.[5]

However, the high overpotential required to drive the hydrogen evolution reaction (HER, $2H^+ + 2e^- \rightarrow H_2$) at the cathode increases the electric energy consumption. Although platinum can evolve hydrogen at near-zero overpotential in acidic media with increased efficiency of water splitting, its high cost and low abundance limit its wide spread use in commercial electrolysis systems.[6] Therefore, it is important to design and develop low-cost, earth-abundant, durable and highly efficient electrocatalysts for HER.

In this context, non-noble active catalysts such as transition metal sulphides,[7,8] selenides,[9,10] phosphides,[11] nitrides,[12] carbides,[13,14] and borides[15], have been widely explored for HER. Based on density functional theory (DFT) calculations, the layered $XS_2$ (where X is Mo or W) are considered among the most efficient HER catalysts, since their free energy of hydrogen adsorption ($\Delta G_{H^*}$) on the $XS_2$ edge is close to thermoneutral (i.e. neither too strong nor too weak), being similar to Pt-group metals.[16,17] The catalytic activity of $XS_2$ is localized at the exposed active edge sites, with their basal planes being semiconducting and catalytically inert. Based on this understanding, strategies have been developed aiming at increasing the concentration of active catalytic sites (both at edges and basal plane) and enhancing the conductivity of basal plane by modulating its electronic structure. Towards this, various exfoliation methods have been developed including ultrasonication[18,19] and grinding[20,21] combined with sequential centrifugation for fine size selection, and exfoliation via lithium intercalation[22]. The first two approaches

a. School of Engineering, Engineering Research Institute, Ulster University, Newtownabbey BT37 0QB, United Kingdom.
b. College of Chemistry and Molecular Engineering, Peking University, Beijing 100871, People's Republic of China.
c. 2-DTech Ltd, Core Technology Facility, 46 Grafton St, Manchester M13 9NT, United Kingdom..
E-mails: lmwx@pku.edu.cn; p.papakonstantinou@ulster.ac.uk

produce pristine XS$_2$ nanosheets (2H-semiconducting phase), where a simultaneous decrease in lateral size and layer number can increase the active electrocatalytic edge sites and enhance the charge transfer in the vertical direction.[20,21] On the other hand, Li-intercalation of bulk crystals followed by reaction with water can produce large quantities of conductive nanosheets (1T metallic phase).[22] However, the reaction is aggressive, bringing a host of safety concerns, with the reaction being extremely air sensitive and by products requiring the need for extensive cleaning.

To circumvent safety drawbacks, an alternative approach to activate the inert basal plane of XS$_2$ involves the introduction of conductive metal oxide phases by in situ electrochemical oxidation.[23,24] In situ electrochemical oxidation of tungsten disulfide (WS$_2$) grown directly on carbon fiber paper at various temperatures induced the surface formation of tungsten trioxide dihydrate (WO$_3$·2H$_2$O) nanoplates, which exhibited enhanced electrochemical HER compared to the unoxidized material.[24] However the mechanism of in situ anodisation on pristine WS$_2$ exfoliated nanosheets and the stability of induced phases during HER accelerated degradation test, have not been investigated before.

High boiling point solvents such as DMF are commonly used in the preparation of inks of exfoliated 2D materials as they provide stable dispersions for long duration. However the aging effect of DMF solvent on WS$_2$ dispersions with well controlled size distributions on HER performance has not been studied before.

Here we report for the synthesis of WS$_2$ nanosheets decorated with scattered nanodots (NSDs) using ionic liquid assisted grinding exfoliation coupled by gradient centrifugation.[20,21] Compared to bulk WS$_2$, WS$_2$ NSDs showed significantly enhanced catalytic activity for HER and a stable performance in acid media (0.5 M H$_2$SO$_4$). However, a deterioration in the catalytic performance was observed for WS$_2$ NSDs dispersions, 1 month old. The aging effect of WS$_2$ nanosized dispersions on the HER behavior has been attributed to the presence organic solvent residues and agglomeration effects. Importantly, the HER activity was recovered by cleaning the electrode surface with acetone. The performance of the pristine NSDs could be further improved by in situ electrochemical oxidation in acid media and the enhancement was found to arise primarily from the formation of amorphous WO$_3$/WS$_2$ heterostructures, which served to provide additional active sites and modify the immediate environment of the WS$_2$ electrocatalyst. However, the durability of the activation induced heterostructures was moderate as revealed by chronoamperometry tests.

# Experimental section

### Preparation of WS$_2$ with different sizes.

WS$_2$ nanosheets/nanodots were prepared by a simple ionic liquid assisted grinding exfoliation method combined with sequential centrifugation steps. WS$_2$ platelets (Sigma-Aldrich, < 2 μm) were ground in an agate mortar grinder (Retsch RM 200; agate mortar and pestle used to ensure no metal contamination during grinding) for 10 hours by mixing with room temperature ionic liquid (1-butyl-3-methylimidazolium hexafluorophosphate, BMIMPF$_6$) in a ratio (1:1) to produce a gelatinous material. The resulting gel was washed three times in a mixture (1:3) of DMF and acetone to remove the room temperature ionic liquid (RTIL). The clean ground product consisted of an assortment of sheets with various sizes and thicknesses, which were subsequently size selected by a sequential centrifugation at 500, 1000, 3000 and 10000 rpm speeds as described in our previous works.[20,21] Isolated products were freeze dried and were labelled as WS$_2$ $X$K, where $X$K represents the centrifugation speed in thousands of rpm.

### Modification of electrodes with WS$_2$ $X$K.

WS$_2$ $X$K were dispersed in the DMF with a concentration of 2 mg ml$^{-1}$. Prior to the modification, a glass carbon electrode (GCE, 3 mm diameter) was polished with a 0.05 μm alumina slurry, and then successively washed ultrasonically with deionized water and ethanol for a few minutes. Then 10 μL of fresh ink was dropped on the pretreated GCE surface resulting in a catalyst loading of 0.283 mg cm$^{-2}$ and dried under an IR lamp. The performances of freshly prepared and 1 month old inks of WS$_2$ 10K were compared under the same catalyst loading on the GCE electrode.

### Electrochemical activation of WS$_2$ 10K.

The WS$_2$ 10K modified electrode was electrochemically activated by applying a constant potential (1 V vs. Ag/AgCl) in 0.5 M H$_2$SO$_4$ for 60 s. Then the electrode was directly used to test HER ability.

### Characterization

Polarization curves were obtained by linear sweep voltammetry at a scan rate of 10 mV s$^{-1}$ on an Autolab, PGSTAT20/FRA system with a protective layer of N$_2$ gas in 0.5 M H$_2$SO$_4$. A high purity graphite rod was used as a counter electrode and an Ag/AgCl (3 M KCl) electrode was used as a reference electrode. Here we used a high purity graphite rod rather than a platinum wire to ensure that no contamination of the working electrode with platinum metal occurred during measurements.[25] In the HER measurements, all potentials were determined with respect to reversible hydrogen electrode (RHE) using the equation of $E$(RHE) = $E$(Ag/AgCl)+0.059 pH+0.210. Working electrodes were pre-conditioned prior to polarization curves by performing cyclic voltammetry at a scan rate of 100 mV s$^{-1}$ for 10 cycles to get stable and reliable HER polarization curves. Electrochemical impedance spectra (EIS) were measured in 0.5 M H$_2$SO$_4$ at -0.20 V (vs. RHE) and 0.1 M KCl-0.01 M phosphate buffer solution (pH = 7.4) containing 5 mM K$_3$Fe(CN)$_6$-K$_4$Fe(CN)$_6$ (1:1) at 0.21 V (vs. Ag/AgCl), respectively. The frequency ranged from 100 000 Hz to 0.1 Hz and the applied AC voltage was 5 mV. Polarization curves from all catalysts were $iR$ corrected, where the R is the ohmic resistance arising from the external resistance of the electrochemical and was measured by EIS in 0.5 M H$_2$SO$_4$.

Scanning electron microscopy (SEM) images were performed on a Quanta 200 3D system. Transmission electron microscopy (TEM) images of as-prepared catalysts were taken on a JEOL JEM-2100F and a FEI TECNAI TF20 TEMs at an accelerating voltage of 200 kV. X-ray diffraction (XRD) patterns were conducted on powdered samples with a Bruker AXS D8 discover with Cu-κα radiation (40 kV, 20 mA, λ = 1.5418 A). UV-vis spectra were acquired from a Perkin Elmer Lamba 35 spectrometer. Spectra were taken in the range of 200 nm to 800 nm in DMF solution. Raman spectra were obtained on micro Raman imaging spectrometer (Thermo Fisher Scientific) with laser excitation wavelength of 532 nm. High resolution X-ray photoelectron spectroscopy (XPS) spectra were obtained on a Kratos AXIS ultra DLD with an Al K$\alpha$ (h$\nu$ = 1486.6 eV) x-ray source. Spectra were fitted to a Shirley background.

## Results and discussion

### A.   WS$_2$ nanosheets/nanodots

WS$_2$ with different lateral sizes and thicknesses were obtained by exfoliation of bulk material combined with sequential centrifugation steps at increasing speeds; a method have been developed and described in a series of our previous papers[18,20,21,26].

The size and morphology of the collected centrifugation products, at different centrifuge speeds, are displayed in Fig. 1. WS$_2$ bulk consists predominantly of large platelets (~ 2 μm), as well as a wide range of smaller flake sizes stacked on larger platelets (Fig. 1a). The lateral size and thickness of the precipitates decreased with gradient centrifugation speed. Large and thick platelets were isolated at a low speed of 500 rpm (Fig. 1b). The smallest and thinnest products were isolated at the highest speed, employed here, of 10K rpm (Fig. 1e and 1f) and they displayed the smoothest morphology among all products due to their ultra-small and thin nature.

High-resolution TEM (HRTEM) was utilized to investigate the atomic structures of the WS$_2$ products and their crystal quality. TEM images from 10K product presented in Fig. 2a and 2b show restacked nanosheets of various lateral dimensions ranging from approximately 100 nm down to only a few nanometers (~ 2 nm) in the form of nanodots. These scattered nanodots spread on the surface of WS$_2$ nanosheets without being uniformly dispersed. HRTEM image in Fig. 2c illustrates clear lattice fringes revealing that the crystalline structure is preserved in the exfoliated products. The d-spacing is 0.27 nm, corresponding to that of (100) planes of 2H-WS$_2$ crystals.[27] The (002) crystal planes, recorded at the folding edge of a nanosheet, enabled the direct observation of the layered structure, revealing that the nanosheets comprised of approximately three atomic layers separated by 0.62 nm, an interlayer spacing value equal to that of 2H-WS$_2$.[28] The selective area electron diffraction (SAED) pattern in Fig. 2d, recorded from the same sample, reveals several sets of patterns with an obvious six-fold symmetry and contains diffraction spots corresponding to the (100) and (110) types of lattice planes.[29]

The effectiveness of the gradient centrifugation procedure in size selecting WS$_2$ products can be further validated by X-ray diffraction analysis as shown in Fig. 3a. WS$_2$ bulk displays an intensive (002) peak as well as some weaker peaks (004), (100), (101), (003), (006), (105), (110) and (112), characteristic of polycrystalline 2H-WS$_2$ according to the JCPDS card no. 08-0237. The interlayer distance calculated according to Bragg's equation, is 0.62 nm agreeing well with HRTEM observation. No substantial change in the (002) peak was observed for WS$_2$ 0.5K due to the exposure of large and highly crystalline platelets as revealed by SEM image of Fig. 1b. However, higher centrifugation speeds caused a progressive broadening and decrease in (002) peak intensity, with WS$_2$ 10K sediment retaining only 6.9% of the bulk intensity. A similar decrease in the diffraction intensity was observed in the sequential centrifugation products from ground MoS$_2$ and graphite.[20,21] The drastic decrease in the intensity of the (002) peak indicates a reduction in the number of aligned planes (layers), and is indicative of highly exfoliated WS$_2$.

The optical properties of the WS$_2$ centrifugation products dispersed in DMF were investigated by UV-vis absorption spectroscopy, shown in Fig. 3b. No detectable peaks were observed in the spectra of WS$_2$ bulk. Bulk transition metal dichalcogenides (TMDs) have indirect band gaps; therefore, absorption peaks do not appear in their UV-vis absorption spectra. In contrast, four characteristic absorption bands (labelled as A, B, C and D) located at 638, 529, 465 and 422 nm were observed in WS$_2$ 10K dispersion, indicative of the presence of a direct band-gap[30]. The excitonic A and B bands at 638 and 529 nm, respectively, correspond to direct gap transitions at the K point in the Brillouin zone.[31,32] The C and D peaks at 465 and 422 nm, are assigned to the direct transitions from the deep valence to the conduction band.[32,33] Collectively, these results significantly diverge from the state-of-the-art exfoliated TMDs by BuLi, in which mixed-phase structures, semiconducting 2H and metallic 1T co-exist[22]. In metallic TMDs, the excitonic bands are basically not featured, while in the current case they are fully developed and resolved demonstrating that the exfoliated nanosheets maintain the expected semiconducting prismatic 2H-WS$_2$ structure.[31] For lower centrifugation speeds, the Mie scattering induced background was substantially reduced and the spectra appeared flatten with less distinct peaks, a characteristic of thicker flakes. In addition, WS$_2$ 10K absorption spectra displayed a blue-shift versus the rest of the products, which is consistent with quantum confinement effects arising from thickness and lateral size reduction.[34]

Raman spectroscopy measurements were also performed to further confirm the exfoliation (Fig. 4a). The Raman spectra were recorded using 532 nm laser. The characteristic Raman peaks at 350.6 and 417.9 cm$^{-1}$ in Fig. 4a for WS$_2$ bulk were clearly observed, assigned to the second order in plane longitudinal acoustic phonons 2LA(M) and out of plane A$_{1g}$ vibrational modes of 2H-phase WS$_2$, respectively.[35] In Fig. 4a and 4b, the frequency trends of the 2LA(M) and A$_{1g}$ modes among different centrifuged WS$_2$ products arise from different layer numbers.[35] The 2LA(M) optical mode blue-shifts with increasing centrifugation speed, which indicates a decrease in layer numbers.[35] However, the A$_{1g}$ mode does not exhibit an obvious trend.[35] The possible reason for this discrepancy could be related to the nature of the samples, as small lateral dimensions of WS$_2$ crystals aggregate during solvent evaporation resulting in stacked formation of various thicknesses, which could have an effect on the A$_{1g}$ mode. Nevertheless, the relative intensity of I$_{2LA}$/I$_{A_{1g}}$ rises gradually with the increase of centrifugation speed, achieving the highest intensity ratio at 10K (Fig. 4c) indicating thinner flakes are present at higher speeds.[8]

The 2H phase of WS$_2$ nanosheets was further confirmed with X-ray photoelectron spectroscopy (XPS). The wide survey scans of bulk and WS$_2$ 10K can be seen in Fig. S1. In addition to intense W and S peaks present in the survey, the clear presence of C and O elements for all catalysts is associated with adventitious impurities, which originate from exposure to atmosphere.[36] Considering the fact that the tungsten signal is sensitive to its oxidation state and coordination geometry, any possible discrepancy of the W $4f_{7/2}$ and W $4f_{5/2}$ core level peaks between the bulk and centrifugation products would be associated with divergence from the 2H structure.[22] The peaks of bulk $WS_2$ (Fig. 5a and 5b) around 32.9 eV and 35.1 eV correspond to W $4f_{7/2}$ and W $4f_{5/2}$, respectively, while the peaks at 162.6 eV and 163.8 eV can be attributed to S $3d_{5/2}$ and S $3d_{3/2}$ orbitals, respectively, which are in good agreement with the binding energies of $W^{4+}$ and $S^{2-}$ in 2H phase of $WS_2$.[37,38] $WS_2$ centrifugation products exhibited nearly the same binding energies for well-defined spin-coupled W and S doublets as those of $WS_2$ crystal. Meanwhile, no obvious signal from the $W^{6+}$ was observed, indicating that there is no obvious oxidation of $WS_2$ nanosheets. Stoichiometric ratios of S to W (1.9 : 1) calculated from the respective integrated areas are close to 2 : 1 demonstrating the expected $WS_2$ phase (Table S1).

The HER catalytic activities of the $WS_2$-particle-modified electrodes were measured in $N_2$-saturated 0.5 M $H_2SO_4$ solution. The prepared $WS_2$ electrodes from the sequential centrifugation products demonstrated progressively improved catalytic performance and gradually declined Tafel slope with the 10K electrode exhibiting the best performance, with an onset potential of -130 mV vs. RHE, an overpotential of 337 mV at the benchmark current density of 10 mA $cm^{-2}$ and a Tafel slope of 80 mV $dec^{-1}$ (Fig. 6a and 6b). The Tafel analysis can reveal the HER mechanism. Tafel slope of 80 mV $dec^{-1}$ implied the HER is ruled by Volmer-Heyrovsky mechanism, with rate limiting step of the electrochemical desorption of an adsorbed hydrogen and $H_3O^+$ to form hydrogen ($H_3O^+$ + $e^-$ + cat.-H → $H_2$ + cat. + $H_2O$).[24,39] It is well known that TMDs can catalyze the hydrogen evolution reaction via active sites on the nanosheets edges. Thus, as smaller nanosheets have more edges, it would be expected to be more effective catalysts. In addition, the ultrathin nature of the nanosheets helps to improve the hopping transport efficiency of the electrons in the vertical direction and thus improves the catalytic performance of the catalyst.[20] $WS_2$ 10K not only shows the best HER performance, but also exhibits an excellent durability. After 1000 scanning cycles from 0 to -0.35 V vs. RHE at a scan rate of 100 mV $s^{-1}$, almost no decrease in performance characteristics was observed (Fig. 6c).

### B. Aging effect of DMF solvent on HER

A number of studies have reported that the most promising solvents for the preparation of stable suspensions of 2D TMDs are highly polar solvents, such as N-methylpyrrolidone (NMP) and N,N'-dimethylformamide (DMF), where the cohesive energies of these solvents are close to interlayer energies of TMDs.[40,41] The effect of various solvents on HER performance with well controlled sizes has not been investigated so far. Moreover, it is rarely mentioned in electrochemical studies, if measurements were conducted from freshly prepared inks or inks stored for weeks or months prior to their use. A recent study on 2 weeks aged dispersions of bulk $MoS_2$ failed to identify any consistent trends most probably due to inconsistencies accrued from different crystal sizes present in the bulk material.[42] Recent work by Gao et al. reported the aging effects of monolayer-$MoS_2$ and $WS_2$ due to adsorption of organic materials from the ambient and the gradual oxidation of TMDs along grain boundaries.[43] Here, we found that the catalytic performance of $WS_2$ 10K dispersion in DMF after 1 month of preparation declined dramatically, with the overpotential at 10 mA $cm^{-2}$ shifting to 483 mV and the Tafel slope increasing to 124 mV $dec^{-1}$, as demonstrated in Fig. 7a and 7b. Similar declining performance issues have also been faced in TMDs electronic devices, prepared from high boiling point solvents such as DMF, where the solvent removal requires high thermal treatment to evaporate the solvent and recover the electrical properties. The observed deteriorated performance can be attributed to two main effects. $WS_2$ nanosheets tend to restack and aggregate after having been stored for some time leading to reduced electroactive surface.[44] During this restacking and aggregation process, organic DMF solvent is both trapped between the $WS_2$ nanosheets and adsorbed onto their surface impeding electron transfer. Interestingly, the performance could be almost recovered to the original value achieved from the fresh dispersion, when the electrode was dipped into acetone and rinsed for 1 min, attaining an onset potential of -136 mV vs. RHE and the overpotential of 364 mV at 10 mA $cm^{-2}$ (Fig. 7a). The Tafel slope also decreased from 124 to 92 mV $dec^{-1}$, close to 80 mV $dec^{-1}$ of fresh $WS_2$ 10K (Fig. 7b), suggesting the rate determining step changed from the reduction of a proton to yield a catalyst-adsorbed H atom ($H_3O^+$ + $e^-$ + cat. → cat.-H + $H_2O$) to electrochemical desorption step ($H_3O^+$ + $e^-$ + cat.-H → $H_2$ + cat. + $H_2O$). Acetone is able to remove DMF residues, which impede electron transfer and cover active sites and itself can evaporate quickly when the washed electrode is dried in the fume hood. This finding is important for practical applications as DMF dispersions can be used for longer times. Except acetone, other organic solvents such as toluene or o-dichlorobenzene (o-DCB), which can be miscible with DMF but have a weak interaction with $WS_2$, also showed a similar "cleaning effect" (Fig. S2**). It should be noted that even for the fresh dispersions, an appreciable improvement on overpotential at 10 mA $cm^{-2}$ was evident for all $WS_2$ sizes after acetone cleaning, shifting positively by 15-75 mV. The enhancement was more apparent for the larger sizes (0.5K and 1K products) (Fig. S3), which indicates that for the larger platelets, DMF is prone to intercalate between the platelets and also be adsorbed on the GCE electrode surface increasing contact resistance. It has been reported that DMF has a low electron affinity (13.6 meV) and as a result does not promote charge transfer[45]. The DMF molecules act as the barrier against charge transfer and thereby increase the contact resistance at the GCE/$WS_2$ interface.

XPS spectroscopy was used to analyze the surface of an electrode modified with aged dispersion in DMF before and after cleaning in order to confirm the effective removal of organic residues with acetone. It is evident from Fig. S4, S5 and 8, that the peak intensities of C 1s, N 1s, O 1s, which are DMF constituents, dramatically decrease, whereas the W 4f and S 2p peak intensities increase after cleaning. These observations indicate that the organic residues could be removed effectively exposing more $WS_2$ active sites. No obvious change could be observed on the binding energies of S 2p and W 4f of aged $WS_2$ 10K before and after cleaning, as illustrated in Fig. 8, suggesting that DMF was not bonded to the $WS_2$ surface.

Electrochemical impedance spectroscopy (EIS) measurements using a reversible redox couple $K_3Fe(CN)_6$ and $K_4Fe(CN)_6$ is a useful probe for comparing the conductivity of different catalysts. In the Fig. S6, the 1 month old $WS_2$ 10K ink exhibits a much larger charge transfer resistance ($R_{ct}$) (~3200 $\Omega$) compared to the fresh ink modified electrode (~1300 $\Omega$), indicating a lower conductivity for 1 month old WS$_2$ 10K. The overall low conductivity inhibited electron transport from the GCE electrode to the catalyst active sites. However, the $R_{ct}$ decreased to about 1400 $\Omega$ after the electrode was rinsed in acetone, a value close to that of the fresh electrode. The above results suggest that the low catalytic performance of old ink electrode is attributed to the existence of organic residues as well as agglomeration of 10K nanosheets, which makes the electrocatalyst less conducting and results in less efficient charge transfer kinetics. The DMF residues serve as "blockages" for electron transport; the WS$_2$ covered with DMF inhibits direct electrical contact between GCE and neighboring nanosheets, providing a reduced electron transport to the active edge sites of the catalyst. After 1000 LSV cycles from 0 to -0.4 V vs. RHE at a scan rate of 100 mV s$^{-1}$, only a slight drop in the overpotential at 10 mA cm$^{-2}$, shifting negatively by 8 mV, was observed, which suggests that the catalyst is stable (Fig. 7c).

## C. Electrochemical anodic activation studies

Electrochemical activation can be considered as an effective approach to modify the electronic structure of basal plane and thus tune its bond energy as well as increase electron transfer kinetics, leading to enhanced catalytic activity.[46,47]

To explore the effect of electrochemical activation potentials, consecutive scans were performed over the range 0 to 1.5 V. As shown in Fig. 7a, the first anodic scan manifested an intense oxidation peak at about 1.0 V that was totally suppressed in the second and third scans. It was further revealed that 1.0 V was the best activation potential as confirmed in Fig. 7b. Since activation potentials higher than 1.0 V had a similar enhancing effect on the HER performance, 1.0 V was selected to avoid the use of higher potentials.

The performance of WS$_2$ 10K was significantly improved through an in situ electrochemical activation (denoted as A-WS$_2$ 10K), where the modified electrode was subjected to a positive voltage (1 V vs. Ag/AgCl) in 0.5 M H$_2$SO$_4$ for 60 s, decreasing the kinetic overpotential by more than 80 mV (WS$_2$ 10K: 255 mV and A-WS$_2$ 10K: 337 mV) at a current density of 10 mA cm$^{-2}$ (Fig. 9a). Furthermore, the A-WS$_2$ 10K exhibited a smaller Tafel slope value of 73 mV dec$^{-1}$, while non-activated WS$_2$ 10K possessed a Tafel slope value of 80 mV dec$^{-1}$ (Fig. 9b). The smaller Tafel slope value of A-WS$_2$ 10K gave rise to a quicker upsurge of current density with the increase of potential. Application of longer activation times (e.g. 300 s) did not improve the HER ability further, indicating that the maximum degree of activation could be achieved within 1 minute. The performance of A-WS$_2$ 10K compared with the non-activated counterpart under same test conditions is shown in the Table S3. The A-WS$_2$ 10K shows much better catalytic capacity than other 2H WS$_2$ on GCE, and even is comparable with 1T metallic WS$_2$ and 2H WS$_2$ on carbon fiber paper or carbon cloth (Table S3). Stability is also a key criterion for evaluating electrocatalysts. Accelerated degradation testing showed that the overpotential for achieving 10 mA cm$^{-2}$ increased by 60 mV after 1000 CV cycles, indicating a moderate stability for A-WS$_2$ 10K (Fig. 9c). Despite this instability, the performance subjected to the durability test was still better than the non-activated WS$_2$ 10K, indicating that the activation was still beneficial.

In order to get an insight into the changes in the electronic structure of WS$_2$ induced by electrochemical activation, XPS studies were carried for the following 2 samples: i) controlled WS$_2$ 10K (denoted as C-WS$_2$ 10K), ii) activated WS$_2$ 10K (A-WS$_2$ 10K). To facilitate a better comparison and avoid any interfering effects from the process in sulfuric acid, as a control we used a WS$_2$ 10K sample, which was immersed into 0.5 M H$_2$SO$_4$ for 60 s, washed by deionized water and dried before being characterized by XPS. Comparison of XPS surveys for the C-WS$_2$ 10K and A-WS$_2$ 10K samples in Fig. S8, discloses a higher O 1s peak intensity for the A-WS$_2$ 10K. The increase of O 1s peak is mainly attributed to the oxidation of W, which was further corroborated by simultaneous changes in W 4f peaks. For O 1s XPS spectra shown in Fig. S9, a binding energy of 530.8 eV in A-WS$_2$ 10K indicates the existence of oxygen bonded to hexavalent tungsten in WO$_3$.[48] The existence of a shoulder at higher binding energies in A-WS$_2$ 10K is related to the presence of non-stoichiometric tungsten oxides (WO$_{3-x}$) and hydroxyl (-OH) groups. The existence of -OH in WO$_{3-x}$ is needed to maintain charge balance, where oxygen vacancies are filled by -OH groups. W 4f and 5p and S 2p XPS spectra of C-WS$_2$ 10K and A-WS$_2$ 10K are shown Fig. 10. The binding energies of W$^{4+}$ 4f$_{7/2}$, W$^{4+}$ 4f$_{5/2}$ and W$^{4+}$ 5p$_{3/2}$ signals are located at 32.9 eV, 35.1 eV and 38.6 eV, respectively, and those of S$^{2-}$ 3d $_{5/2}$ and S$^{2-}$ 3d $_{3/2}$ are at 162.6 eV and 163.8 eV respectively, all of which are representative of 2H-WS$_2$ phase.[24] The peaks located at 169.1 (= 2p$_{3/2}$) eV and 170.2 (= 2p$_{1/2}$) eV can be attributed to +6 orbitals, which originated from the oxidation of divalent sulfide ion (S$^{2-}$) and sulfuric acid.[37,49] There is no obvious change in S region between the C-WS$_2$ 10K and A-WS$_2$ 10K, implying the S is almost kept unchanged during the activation process. The signals with peak binding energies at 36.3 eV and 38.6 eV are attributed to W$^{6+}$ 4f$_{7/2}$ and W$^{6+}$ 4f$_{5/2}$, which are indicative of the presence of WO$_3$ constituent.[24,37] The small amount of WO$_3$ constituent existed in C-WS$_2$ 10K due to the slight oxidation of WS$_2$ surface. However, the atomic ratio of the ratio of W(VI) to W(IV) in A-WS$_2$ 10K increased, with S to W decreasing, compared to C-WS$_2$ 10K (Table S2), suggesting that part of WS$_2$ was oxidized to WO$_3$. Overall, all the XPS results indicate the formation of mainly tungsten trioxide oxide species on the surface of WS$_2$, even though the presence of a small quantity of sub-stoichiometric WO$_{3-x}$ cannot be ruled out. This presence of tungsten trioxide oxide species could be further confirmed by Raman analysis shown in the Fig. S10. The Raman spectrum of A-WS$_2$ 10K contains the signature vibrational peaks of O-W$^{6+}$-O stretching modes (new peaks at 696.1 and 804.8 cm$^{-1}$ respectively) of the bridging oxygen atoms in WO$_3$.[50,51] The low signal counts are indicative of small film thickness and low crystallinity, which are both intuitively correct bearing in mind the short activation time (60 s) and the room temperature treatment (absence of any high processing temperatures) used here. Therefore, it is reasonable to assume that the formation of thin tungsten trioxide induced by electrochemical activation of WS$_2$ at positive potential is responsible for the enhanced HER activity. However, it is well known that bulk WO$_3$ itself, has poor HER and low electrical conductivity and therefore the improvement cannot be directly rationalized.[52,53]

To gain insight into the possible electronic changes induced by the electrochemical activation, electrochemical impedance studies were carried out using reversible $K_3Fe(CN)_6/K_4Fe(CN)_6$ probes, as shown in Fig. S11. After anodic activation, the charge transfer resistance ($R_{ct}$) increased from 1300 $\Omega$ ($WS_2$ 10K) to 3600 $\Omega$ ($A-WS_2$ 10K), indicating a decrease in conductivity. Surprisingly, we found that the conductivity increased, with $R_{ct}$ decreasing to about 1850 $\Omega$, just after performing HER measurement on $A-WS_2$ 10K. Such decrease in conductivity obviously indicates that the application of negative voltage on the $A-WS_2$ 10K modulates its electronic structure and effectively facilitates a transition to a more metallic state.

Here we propose the following mechanism in order to explain the high electrocatalytic activity of the amorphous $WO_3/WS_2$ heterostructures and the observed change in its electrical conductivity. Under cathodic polarization, during the HER scan, the amorphous $WO_3$ ($a-WO_3$) undergoes protonation, where protons ($H^+$) from the acid solution, and electrons injected from the underneath electrode are intercalated in the oxide forming tungsten bronzes ($a-H_xWO_3$).[54] Upon subsequent application of an anodic bias, electrons and protons are extracted and the tungsten bronzes ($a-H_xWO_3$) are oxidized into the original $a-WO_3$.[55] This charge/discharge process, which is the basic mechanism of electrochromic devices is indicated in reaction (1).

$$a-WO_3 + xe^- + xH^+ \rightarrow a-H_xWO_3 \qquad (1)$$

To confirm this process, CV curves were performed on activated and non-activated $WS_2$ 10K samples (Fig. S12). Clear reversible cathodic and anodic current peaks for $A-WS_2$ 10K were observed near -0.2 V (vs. Ag/AgCl), which are associated with reversible electrochemical injection of both protons and electrons into the host oxide, testifying the validity of the above hypothesis.

It has been reported[55] that at negative potentials the intercalation process in $WO_3$ leads to electron reduction from a $W^{6+}$ to $W^{5+}$ states giving rise to sub-stoichiometric $WO_{3-x}$, which possesses metallic conductivity due to the presence of oxygen vacancies. Therefore, based on previous experimental evidence, it is reasonable to assume that a transition from semiconducting ($WO_3$) to metallic phase ($WO_{3-x}$) is occurring during the intercalation process, which is in agreement with the observed fall in impedance at the end of the HER test. Notably, vacancies in sub-stoichiometric $WO_3$ are active sites for HER[51,52,56], and give rise to a high electrocatalytic activity in these oxides. Overall, the intercalated $a-WO_3$ ($a-H_xWO_3$) serves not only to provide additional active sites but also modifies the immediate environment of the $WS_2$ electrocatalyst, rendering it more favourable for local proton delivery and/or transport to the active edge sites of $WS_2$.

To understand the stability of $a-WO_3$ created by the activation process, $A-WS_2$ 10K subjected to HER at a constant potential of -0.30 V vs. RHE (chronoamperometric measurements) for 0.5 h and 2h (labelled as $A-WS_2$-0.5h and $A-WS_2$-2h respectively) were characterized by XPS (Fig. 10). It should be noted that chronoamperometric rather than cyclic voltammetry (CV) test were used, in order to avoid potential re-oxiditation of the surfaces. No obvious change in S region was observed. However, the peak intensity of $W^{6+}$ decreased gradually with the HER test time increasing, indicating the decrease of $a-WO_3$ could result in the drop of the catalytic performance. This decay may be associated with partial dissolution of $a-WO_3$ in agreements with previous reports[54] which claim poor chemical stability and dissolution of even in mildly acidic electrolytes. Based on this, we hypothesized that if the positive volt was applied again, the HER performance might return back. However, the no enhancement could be achieved.

To provide a direct comparison, we also obtained XPS spectra from non-activated $WS_2$ subjected to 2 hours HER test (labelled as $C-WS_2$-2h), shown in Fig. S13. No obvious change was observed for the $WS_2$-2h, demonstrating that the changes in $A-WS_2$-0.5h and $A-WS_2$-2h are attributed to a reduction of $a-WO_3$ and do not originate from changes in $WS_2$ crystal. The atomic ratios of S to W and W(VI) to W(IV) of different materials are summarized in Table S2. For $C-WS_2$ 10K and $C-WS_2$-2h, the ratios of S to W and W(VI) to W(IV) are almost kept the same. The $A-WS_2$ 10K shows the lowest ratio of S to W and highest ratio of W(VI) to W(IV), because of the formation of $a-WO_3$. The ratio of S to W increased and the ratio of W(VI) to W(IV) decreased gradually in the $A-WS_2$-0.5h and $A-WS_2$-2h samples, indicating the $a-WO_3$ was reduced during the HER measurement.

The electrode kinetics under catalytic HER operating conditions (0.5 M $H_2SO_4$) were investigated by EIS, by applying a voltage at -0.2 V, close to the onset potential. In the Fig. S14, $A-WS_2$ 10K showed the lowest $R_{ct}$ (~460 $\Omega$) compared to $WS_2$ 10K (~2900 $\Omega$) corresponding to the fast shuttling of electrons during HER. After 2h galvanostatic test, the $R_{ct}$ increased to about 2300 $\Omega$. The trend is consistent with the HER polarization curves.

Electrochemically active surface area (ECSA) is an important factor to explain the changes of the intrinsic activities. It can be conducted by measurement of the double-layer capacitance ($C_{dl}$) in a potential region with no faradaic response[57]. The ECSA was estimated from the ratio of the measured double layer capacitance with respect to the specific capacitance of an atomically smooth $WS_2$ material (~60 $\mu F/cm^2$).

$$ECSA = \frac{C_{dl}}{C_S} \qquad (1)$$

It should be mentioned that the above method is suitable for electrodes consisting of conductive materials, which could lead to error in the ECSA determination for semiconducting layers, such as $2H-WS_2$, where an increase in active surface area does not necessarily translate into an increase in the double layer capacitance.

To evaluate the double-layer capacitance, cyclic voltammogram scans were acquired in a non-Faradaic region between 0.16 V to 0.36 V (vs. RHE) at various scan rates (Fig. S15). The $C_{dl}$ was calculated from the slope of the straight line ($j = vC_{dl}$), when the charge current density at a particular potential is plotted against scan rate (Fig. S15).

Using the equation (1) we have obtained values of 17.2, 74.3 and 15.9 for the ECSA of $WS_2$ 10K, activated $WS_2$ ($A-WS_2$ 10K) and activated $WS_2$ after performing HER tests for 2 hours ($A-WS_2$-2h) respectively. The ECSA of $A-WS_2$ 10K was 4.3 times larger than that of $WS_2$ 10K, while the ECSA of $A-WS_2$-2h is similar to the starting $WS_2$ 10K. These results confirm further the increase of active sites after electrochemical anodisation and their progressive reduction after HER tests, which corroborate well with the XPS and electrochemical impedance results.

From all the above investigations, it is can be recognized that the immediate contact of the $a-WO_3$ with the $WS_2$ in the a-

WO$_3$/WS$_2$ heterostructures, facilitates proton delivery and electron transport for enabling efficient hydrogen evolution reaction. Although the stability of the hybrid catalyst is not exceptionally good, our study has helped to understand the fundamental mechanisms of HER activity in these heterostructures. Moreover, our study paves the way for the design of different types of more robust tungsten oxides/tungsten disulfide heterostructures, which can improve the HER by utilizing the synergistic effects. For example, by fabricating crystalline WO$_{3-x}$/WS$_2$ heterostructures the stability of HER is expected to improve. It is well known that crystalline WO$_{3-x}$ are stable in acidic electrolyte solutions in contrast to amorphous WO$_3$ films, which exhibit a high dissolution rate in acids. Atomic layer deposition is a suitable technique for the deposition of ultrathin WO$_{3-x}$/WS$_2$ layers.

## Conclusions

In summary this work has three main thrusts. First, we have demonstrated that bulk tungsten disulfide (WS$_2$) can be exfoliated to nanosheets/nanodots with a thickness of about three layers and lateral size dimensions in the range 100 to 2 nm using ionic liquid assisted grinding exfoliation of bulk platelets followed by gradient centrifugation. We showed that the catalytic H$_2$ evolution could be drastically improved by a simultaneous decrease in the layer number and the lateral size, so that the number of active edge sites was increased and efficient electron transport to the catalyst active sites was facilitated. The obtained nanosheets/nanodots exhibit an onset potential as small as -130 mV vs. RHE, an overpotential of 337 mV at a current density of 10 mA cm$^{-2}$, a Tafel slope of 80 mV dec$^{-1}$ and a good long-term stability. A second thrust is the investigation of the effect of aging time of DMF dispersion on HER catalytic ability. The catalytic performance of WS$_2$ declined dramatically after 1 month due to the aggregation of nanosheets and adsorption of DMF on the catalyst surface. We showed that the catalytic activity could be recovered to a level close to the fresh dispersions by rinsing the electrode in the acetone. It is worth noticing that DMF is prone to adsorb and intercalate between the larger platelets/nanosheets obstructing electron transport, so even for electrodes prepared from fresh DMF dispersions, their activities could still be improved after rinsing them in the acetone. The third thrust of this work is the investigation of electrochemical anodisation of WS$_2$ nanosheets/nanodots on HER activity and their mechanisms. On the basis of XPS, Raman and EIS results, the HER improvement upon anodic activation strongly relied on promoting the formation of an amorphous WO$_3$ layer on the surface of WS$_2$ catalyst. The a-WO3 under the influence of negative voltages served to provide additional active sites and modify the immediate environment of the WS$_2$ electrocatalyst, rendering it more favourable for proton delivery. However, the HER stability of these a-WO$_3$/WS$_2$ heterostructures proved mediocre since the tungsten trioxide could be reduced upon accelerated degradation studies and galvanostatic testing. Nevertheless, the performance of activated WS$_2$ nanosheets/nanodots was still competitive, when compared with the non-activated counterparts under same test conditions. The important findings of the current work are concerned with the role of size selected exfoliation, organic solvents and activation on the catalytic performance of WS$_2$, all of which should be taken into consideration for their practical implementation in water splitting devices.

## Acknowledgements

W. L. thanks the China Scholarship Council (CSC) for support. M. L. and P.P. acknowledge financial support from the National Natural Science Foundation of China (No. 21475003 and 21675003) and British Council (Ref: 216182787) respectively.

## Competing financial interests

The authors declare no competing financial interests.

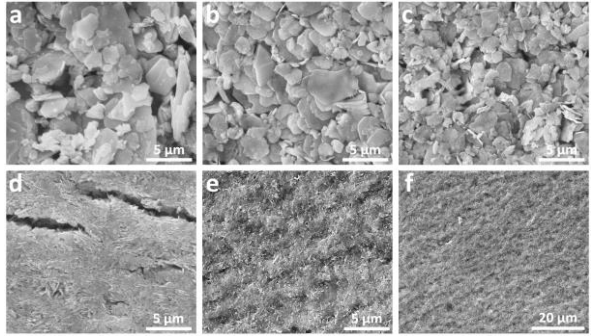

**Fig. 1** SEM images of isolated sediments: (a) WS$_2$ bulk, (b) WS$_2$ 0.5K, (c) WS$_2$ 1K, (d) WS$_2$ 3K, (e) WS$_2$ 10K and (f) WS$_2$ 10K at lower magnification.

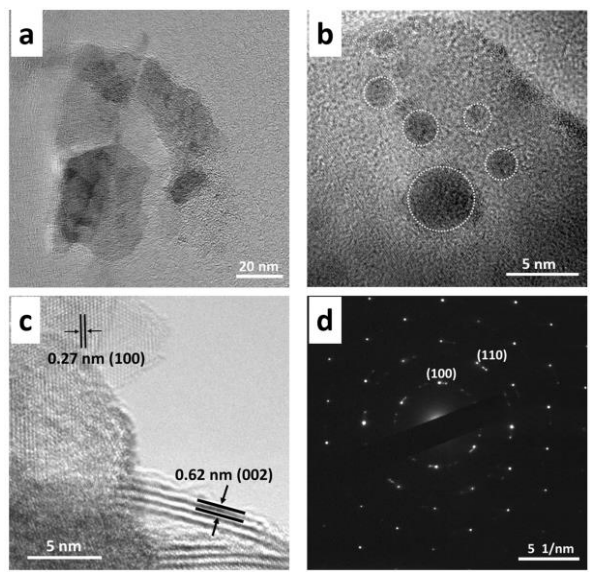

**Fig. 2** (a) and (b) TEM images of WS$_2$ 10K centrifugation products. Dotted circles in (b) represent nanodots. (c) HRTEM image of WS$_2$ 10K centrifugation products. (d) SAED pattern of WS$_2$ 10K centrifugation products.

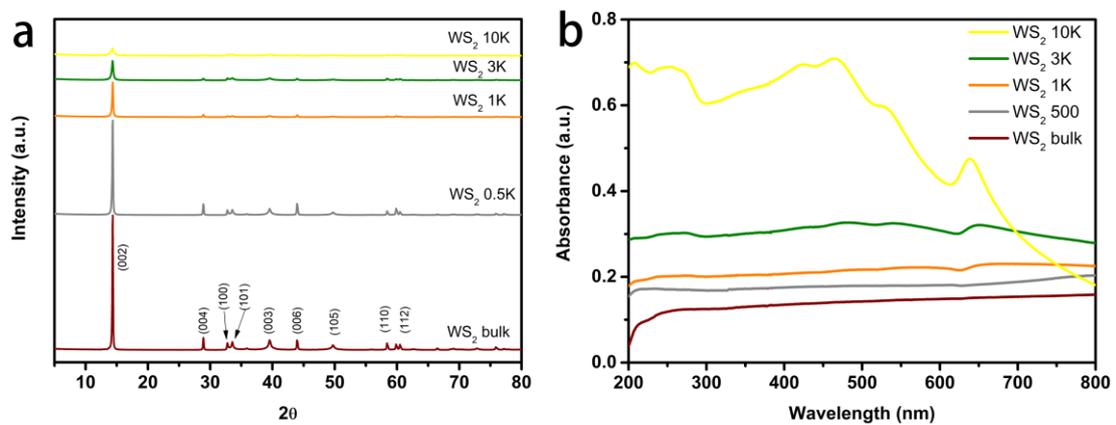

**Fig. 3** (a) XRD patterns of WS$_2$ bulk and 0.5K, 1K, 3K and 10K centrifugation products. (b) UV-vis absorption spectrum for WS$_2$ bulk and all centrifugation products dispersed in DMF.

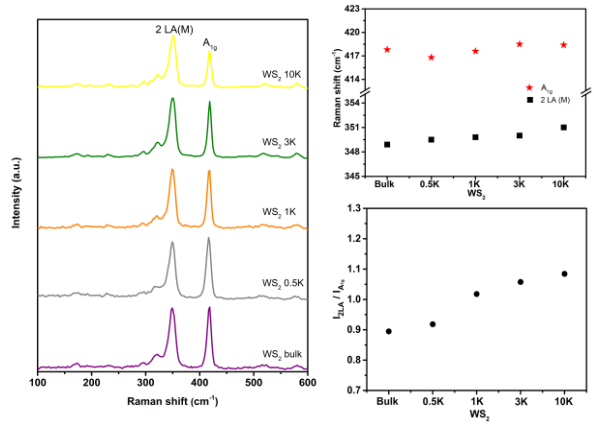

**Fig. 4** Raman characterisation of WS$_2$ materials. a) Raman spectra of WS$_2$ bulk, 0.5 K, 1K, 3K and 10K centrifugation products. b) Frequency peak position of 2LA(M) and A$_{1g}$ modes as a function of centrifugation speeds. c) Intensity ratios of I$_{2LA}$/I$_{A1g}$ as function of centrifugation speeds.

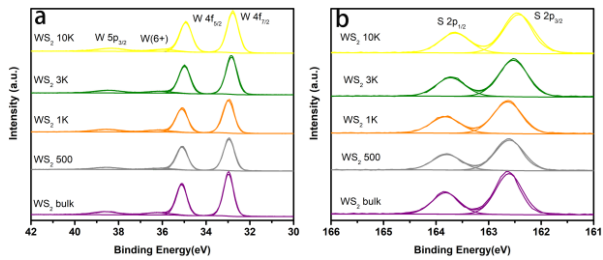

**Fig. 5** High-resolution (a) W 4f and 5p and (b) S 2p core level XPS spectra of WS$_2$ bulk and 0.5K, 1K, 3K and 10K centrifugation products.

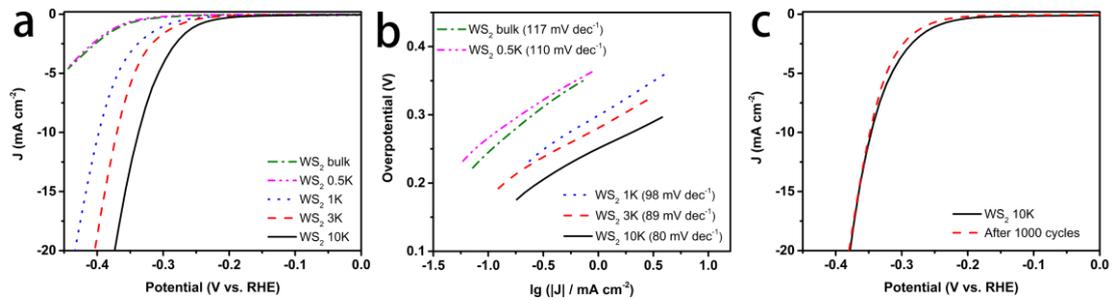

**Fig. 6** (a) Polarization curves and (b) Tafel slopes of WS$_2$ bulk and 0.5K, 1K, 3K, 10K centrifugation products. (c) Polarization curves of WS$_2$ 10K before and after 1000 cycles scan.

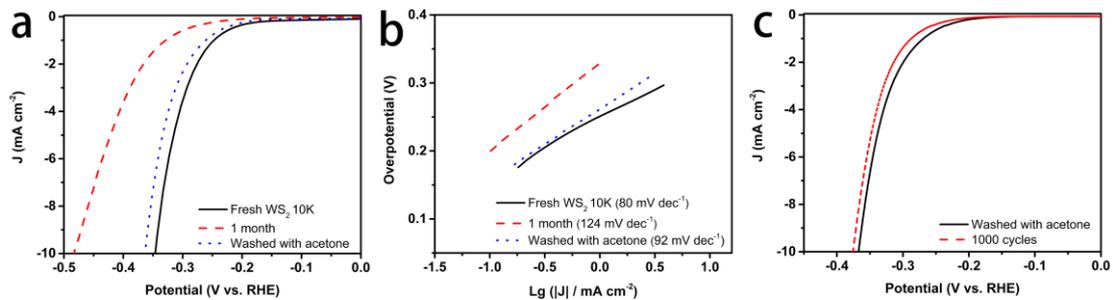

**Fig. 7** (a) Polarization curves and (b) Tafel slopes of the electrodes modified with fresh WS$_2$ 10K and WS$_2$ 10K

dispersions prepared for 1 month (old WS$_2$ 10K), and the old WS$_2$ 10K modified electrode washed with acetone. (c) Polarization curves of old WS$_2$ 10K washed with acetone before and after 1000 cycles scan.

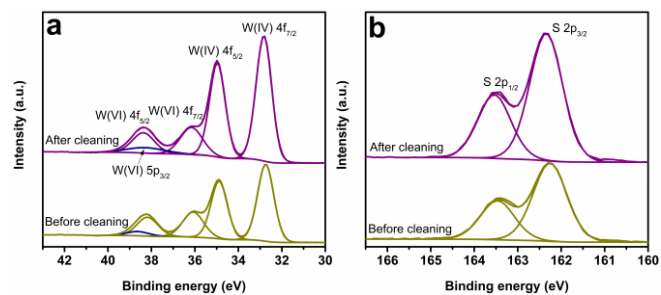

**Fig. 8** High-resolution (a) W 4f and 5p and (b) S 2p core level XPS spectra from a GCE electrode modified with 1 month old WS$_2$ 10K dispersion before and after cleaning in acetone.

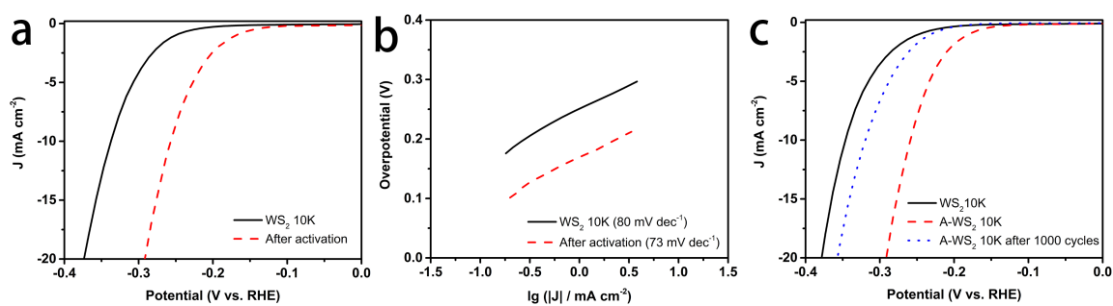

**Fig. 9** (a) Polarization curves and (b) Tafel slopes of WS$_2$ 10K before and after activation. (c) Polarization curves of WS$_2$ 10K and A-WS$_2$ 10K before and after 1000 cycles scan.

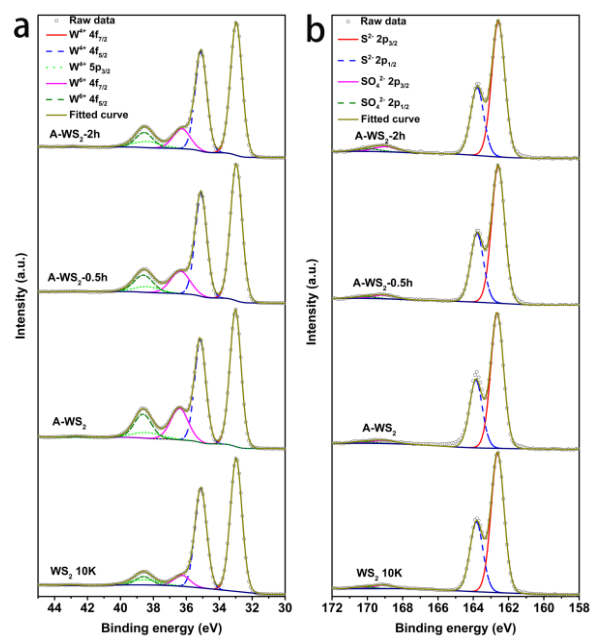

**Fig. 10** High-resolution (a) W 4f and 5p and (b) S 2p core level XPS spectra of C-WS$_2$ 10K, A-WS$_2$ 10K, A-WS$_2$-0.5h and A-WS$_2$-2h.

# Supporting Information

## The Effects of Exfoliation, Organic Solvents and Anodic Activation on Catalytic Hydrogen Evolution Reaction of Tungsten Disulfide

Wanglian Liu,[a, b] John Benson,[c] Craig Dawson,[c] Andrew Strudwick,[c] Arun Prakash Aranga Raju,[c] Yisong Han, [a] Meixian Li,*[, b] and Pagona Papakonstantinou*[, a]

[a] School of Engineering, Engineering Research Institute, Ulster University, Newtownabbey BT37 0QB, United Kingdom.

[b] College of Chemistry and Molecular Engineering, Peking University, Beijing 100871, People's Republic of China.

[c] 2-DTech Ltd, Core Technology Facility, 46 Grafton St, Manchester M13 9NT, United Kingdom.

*Corresponding authors' e- mails: lmwx@pku.edu.cn; p.ppakonstantinou@ulster.ac.uk

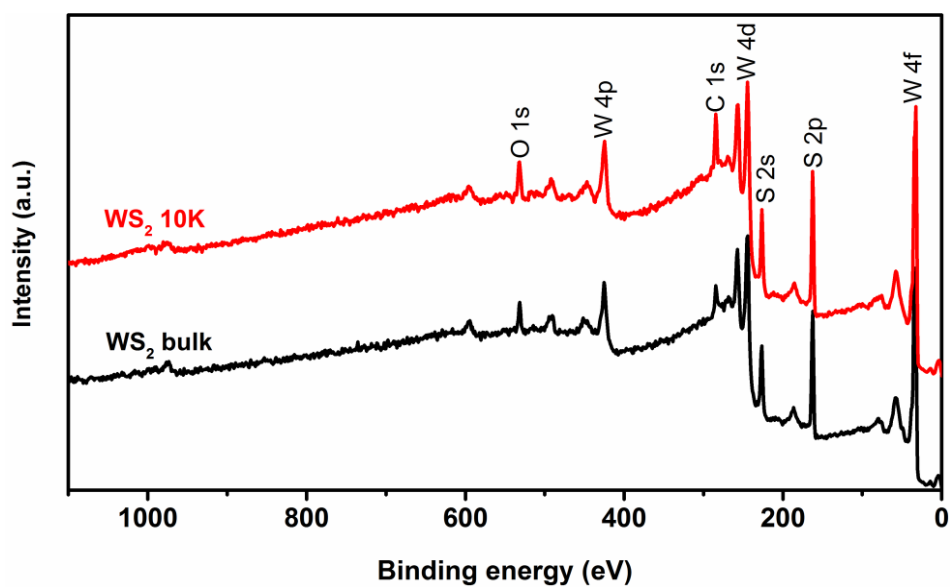

**Fig. S1** XPS surveys of WS$_2$ bulk and WS$_2$ 10K.

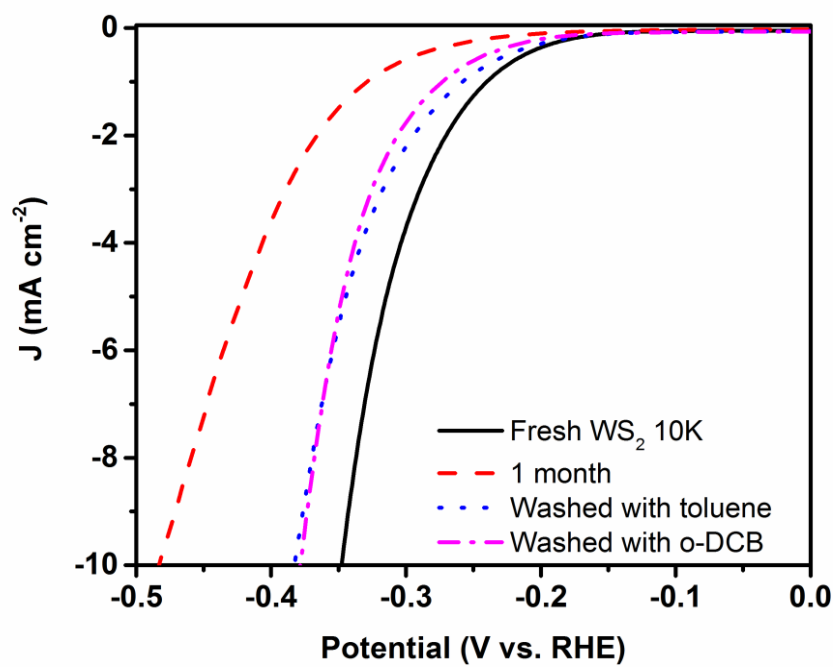

**Fig. S2** Polarization curves of electrodes modified with i) freshly prepared WS$_2$ 10K dispersion; ii) WS$_2$ 10K dispersion 1 month old (1 month); iii) WS$_2$ 10K dispersion 1 month old, washed with toluene and iv) WS$_2$ 10K dispersion 1 month old, washed with o-dichlorobenzene (o-DCB).

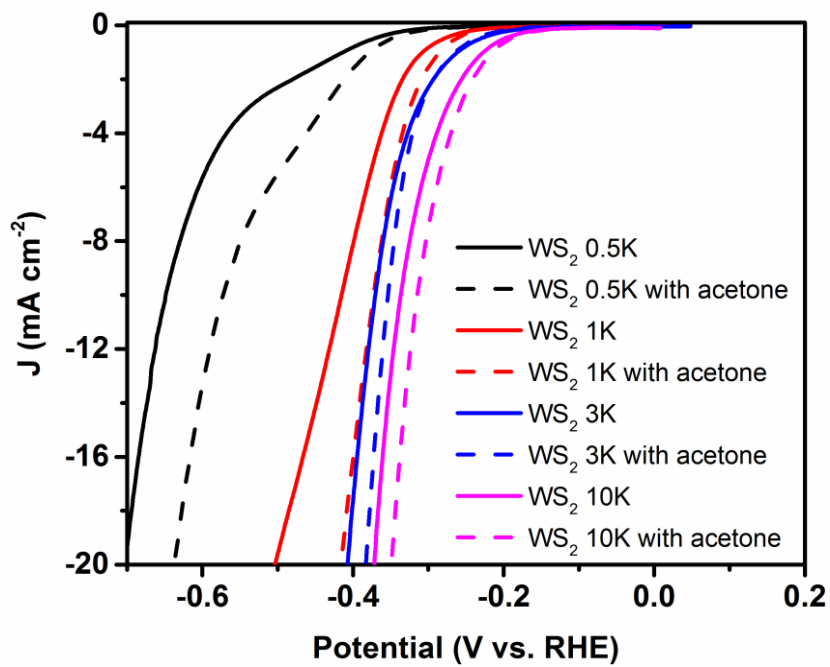

**Fig. S3** Polarization curves of the electrodes modified with fresh 0.5K, 1K, 3K and 10K dispersions before (full lines) and after (dashed lines) washing with acetone.

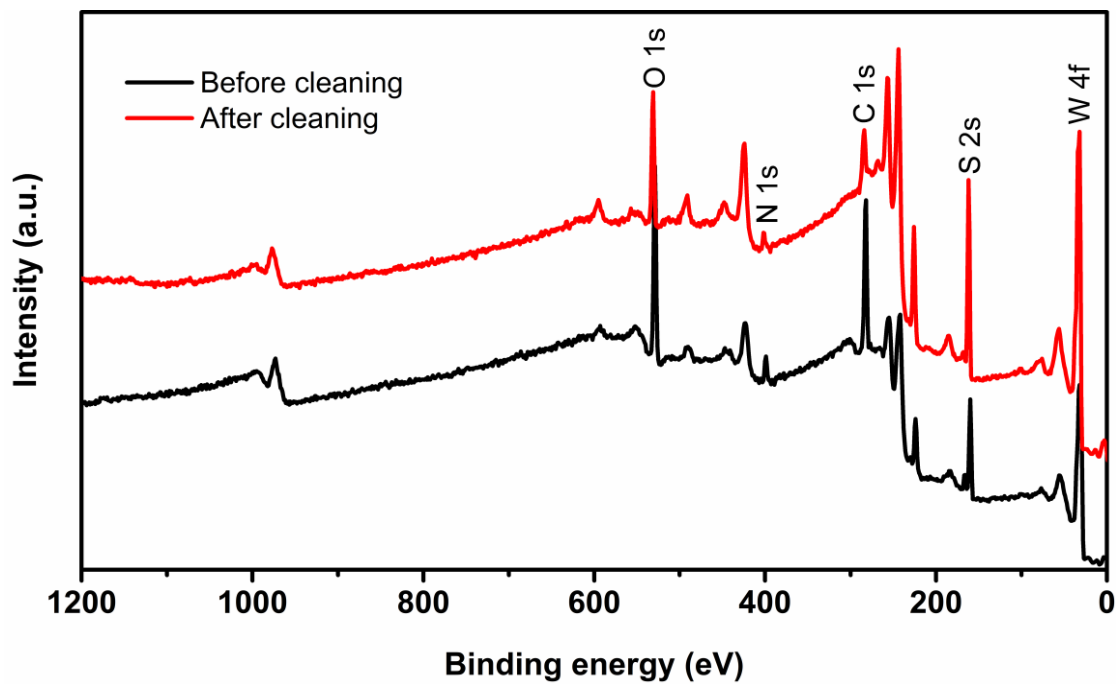

**Fig. S4** XPS surveys of a GCE electrode modified with 1 month old WS$_2$ 10K dispersion before and after cleaning with acetone.

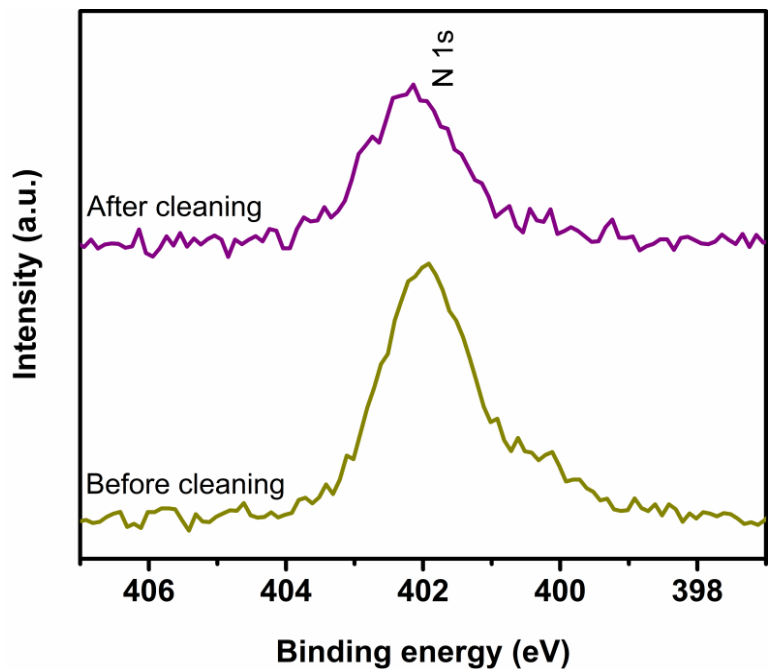

**Fig. S5** High-resolution N1s core level XPS spectra from a GCE electrode modified with 1 month old WS$_2$ 10K dispersion before and after cleaning in acetone.

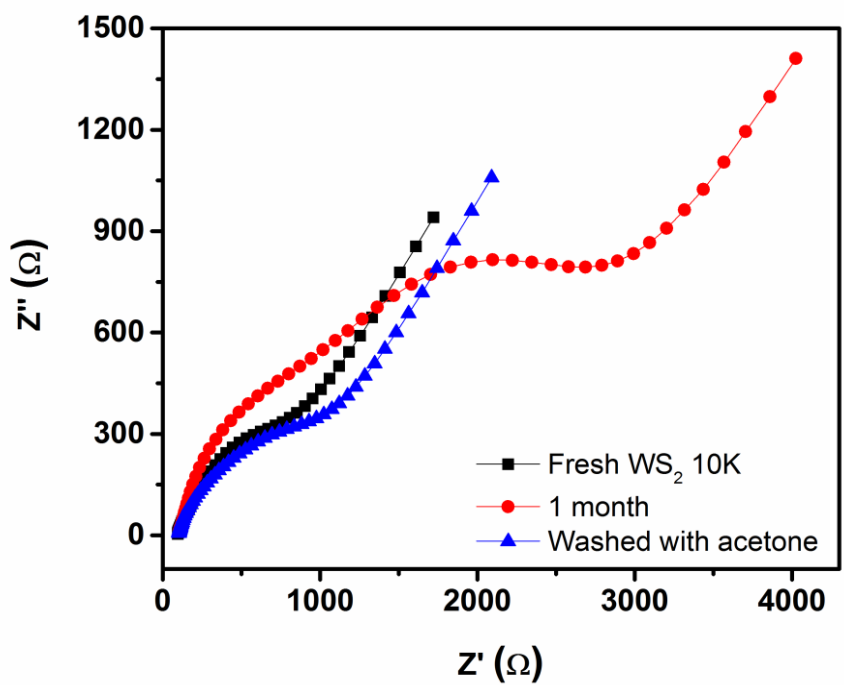

**Fig. S6** Electrochemical impedance spectra of the electrodes modified with i) freshly prepared WS$_2$ 10K dispersion; ii) WS$_2$ 10K dispersion 1 month old and iii) WS$_2$ 10K dispersion 1 month old, washed with acetone. The measurements were conducted in 0.1 M KCl-0.01 M phosphate buffer solution (pH = 7.4) containing 5 mM K$_3$Fe(CN)$_6$-K$_4$Fe(CN)$_6$ (1:1).

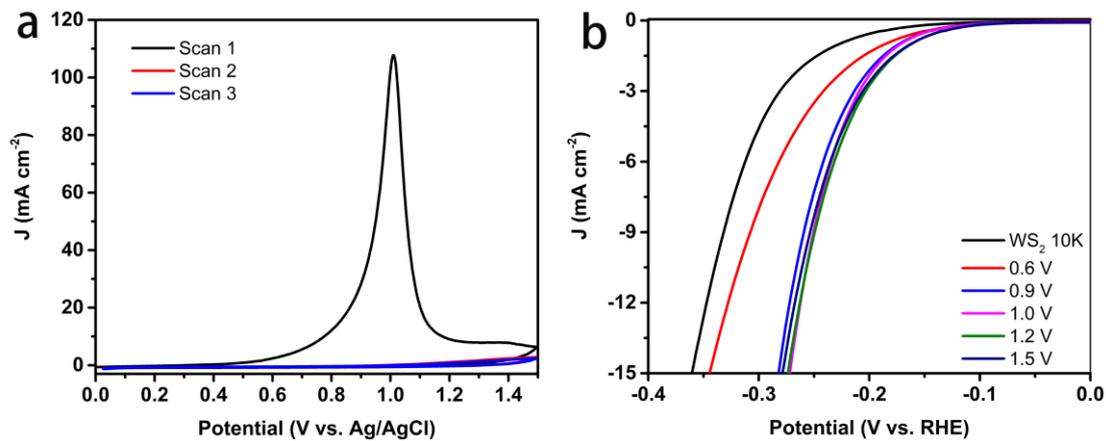

**Fig. S7** a) Cyclic voltammetry scans of a WS$_2$ 10K modified electrode performed between 0 and 1.5 V (vs. Ag/AgCl) in 0.5 M H$_2$SO$_4$. b) Polarization curves of WS$_2$ 10K modified electrodes before and after activation at different potentials in 0.5 M H$_2$SO$_4$.

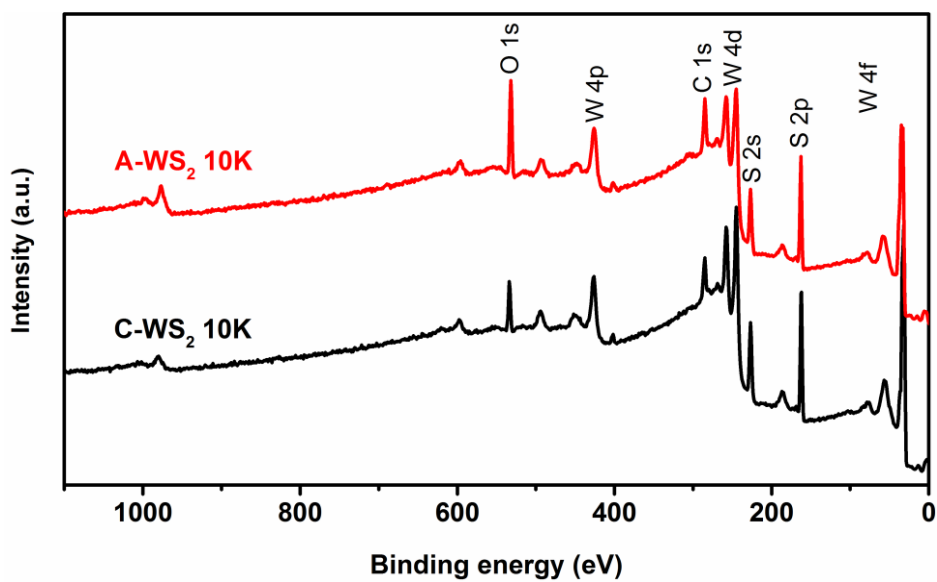

**Fig. S8** XPS surveys of WS$_2$ 10K before (C-WS$_2$ 10K) and after (A-WS$_2$ 10K) activation.

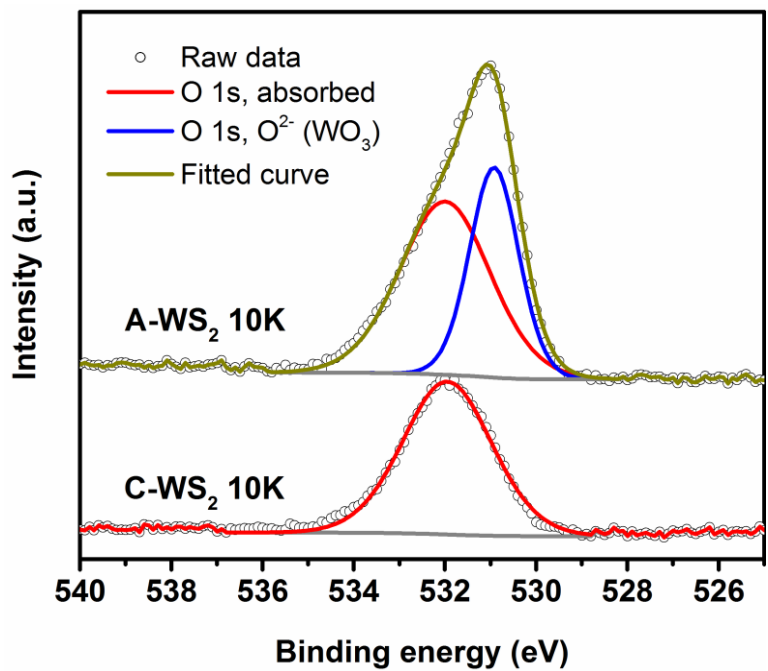

**Fig. S9** High resolution O 1s core level XPS spectra of WS$_2$ 10K before (C-WS$_2$ 10K) and after (A-WS$_2$ 10K) activation.

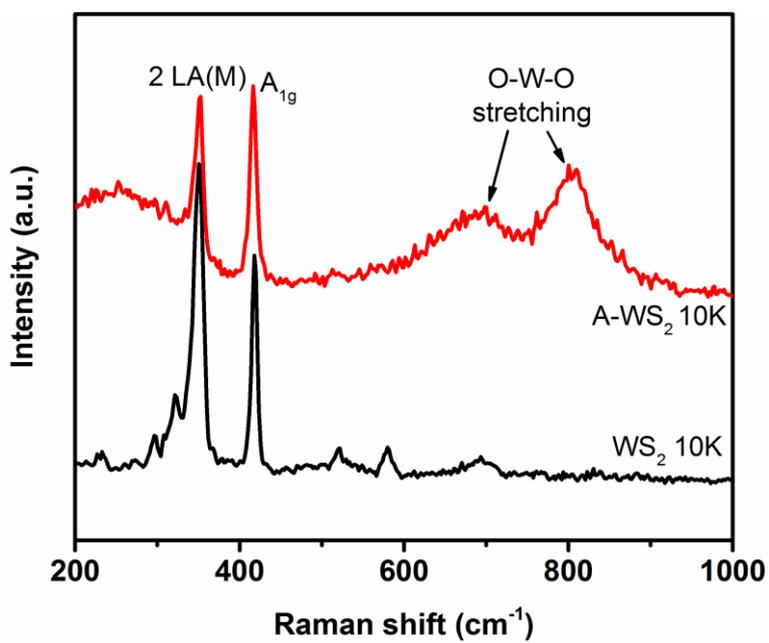

**Fig. S10** Raman spectra of WS$_2$ 10K before (WS$_2$ 10K) and after (A-WS$_2$ 10K) activation.

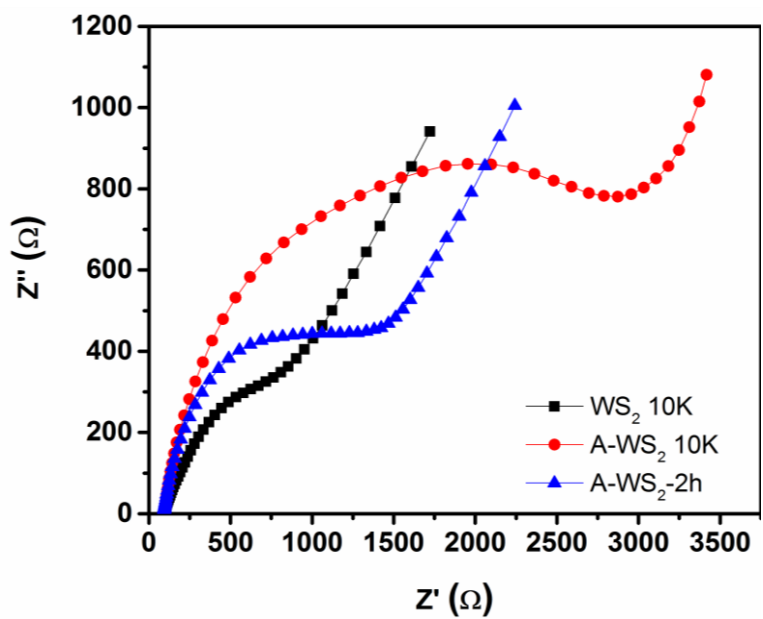

**Fig. S11** Electrochemical impedance spectra of a WS$_2$ 10K electrode: i) before activation (WS$_2$ 10K); ii) after activation (A-WS$_2$ 10K) and iii) after activation and 2 hours HER test (A-WS$_2$-2h). EIS was conducted in 0.1 M KCl-0.01 M phosphate buffer solution (pH = 7.4) containing 5 mM K$_3$Fe(CN)$_6$-K$_4$Fe(CN)$_6$ (1:1).

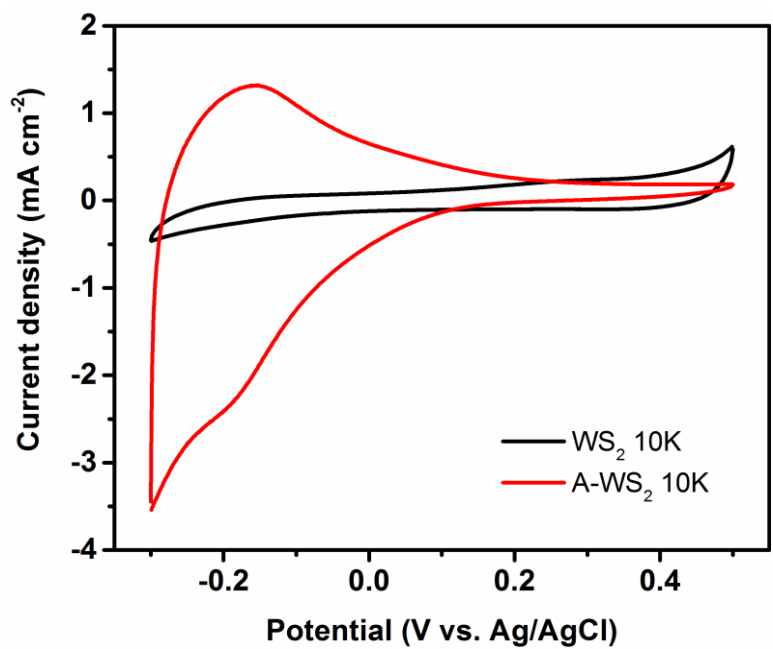

**Fig. S12** Cyclic voltammograms performed between 0.5 and -0.3V (vs. Ag/AgCl) with a scan rate of 100 mV s$^{-1}$ in 0.5 M H$_2$SO$_4$ for i) WS$_2$ 10K and ii) activated WS$_2$ 10K (A-WS$_2$ 10K), respectively.

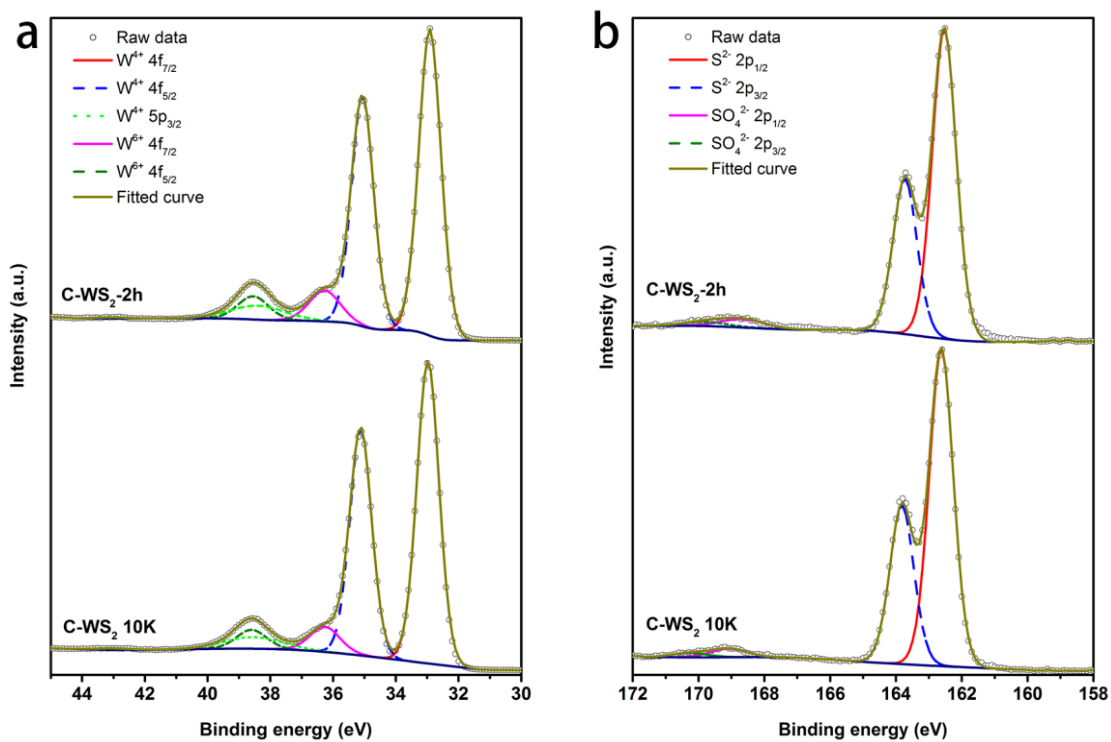

**Fig. S13** High resolution (a) W 4f and 5p and (b) S 2p core level XPS spectra of controlled WS$_2$ 10K before (C-WS$_2$ 10K) and after (C-WS$_2$-2h) 2h HER test.

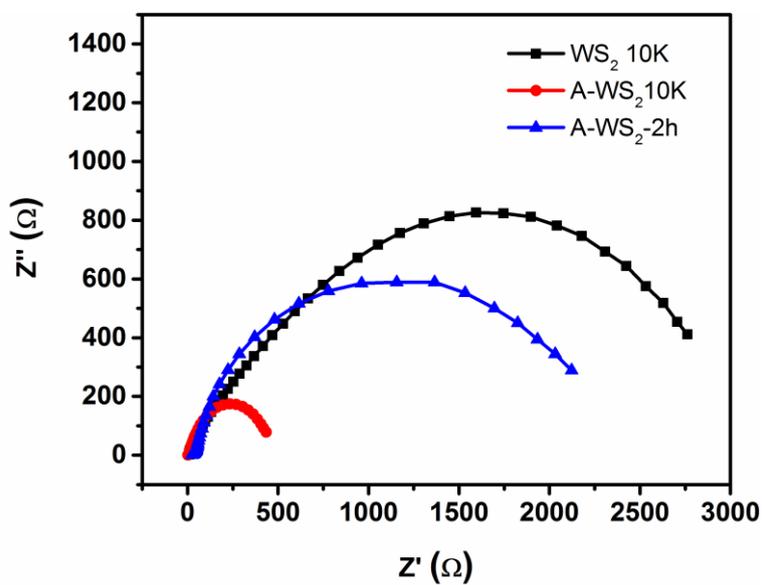

**Fig. S14** Electrochemical impedance spectra of i) WS$_2$ 10K; ii) activated WS$_2$ 10K (A-WS$_2$ 10K) and iii) activated WS$_2$ 10K subjected to 2 hours HER test (A-WS$_2$-2h) in 0.5 M H$_2$SO$_4$.

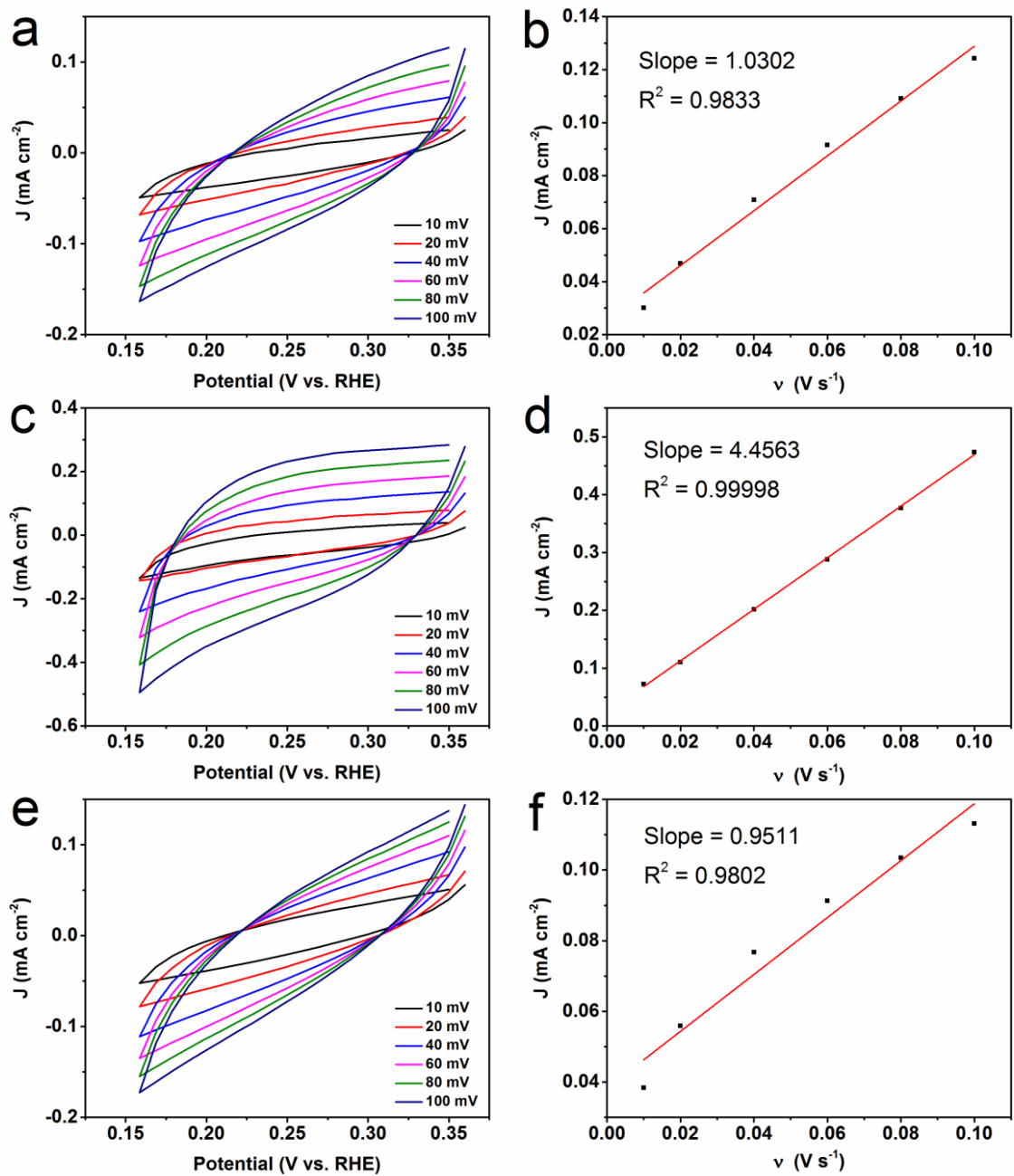

**Fig. S15.** Cyclic voltammogram scans of (a) WS$_2$ 10K, (c) a-WS$_2$ 10K and (e) a-WS$_2$-2h at different scan rates (0.1, 0.08, 0.06, 0.04, 0.02 and 0.01 V s$^{-1}$) and the corresponding capacitive current density measured at 0.25 V (vs. RHE) plotted as a function of scan rate (b, d, f). The average value of the slope was determined as the double-layer capacitance (C$_{dl}$) of each catalyst. The calculated C$_{dl}$ of WS$_2$ 10K, A-WS$_2$ 10K and A-WS$_2$-2h are 1.03, 4.46 and 0.95 mF cm$^{-2}$, respectively.

**Table S1** XPS data of composition of WS$_2$ bulk and 500, 1K, 3K and 10K centrifugation products.

| | WS$_2$ bulk | WS$_2$ 0.5K | WS$_2$ 1K | WS$_2$ 3K | WS$_2$ 10K |
|---|---|---|---|---|---|
| S (%At conc) | 65.5 | 65.2 | 65.2 | 65.4 | 66.0 |
| W (%At conc) | 34.5 | 34.1 | 34.8 | 34.6 | 34.0 |
| Ratio of S to W | 1.9 : 1 | 1.9 : 1 | 1.9 : 1 | 1.9 : 1 | 1.9 : 1 |

**Table S2** The atomic ratios of S to W and W(VI) to W(IV) of controlled $WS_2$ 10K (C-$WS_2$ 10K), activated $WS_2$ 10K (A-$WS_2$ 10K), activated $WS_2$ 10K subjected to 0.5 and 2 hours HER test (A-$WS_2$-0.5h and A-$WS_2$-2h) and controlled $WS_2$ 10K subjected to 2 hours HER test (C-$WS_2$-2h).

| | C-$WS_2$ 10K | A-$WS_2$ 10K | A-$WS_2$-0.5h | A-$WS_2$-2h | C-$WS_2$-2h |
|---|---|---|---|---|---|
| Ratio of S to W | 1.9 : 1 | 1.6 : 1 | 1.7 : 1 | 1.8 : 1 | 1.9 : 1 |
| Ratio of W(VI) to W(IV) | 0.1 : 1 | 0.4 : 1 | 0.3 : 1 | 0.2 : 1 | 0.1 : 1 |

**Table S3** Comparison of the electrocatalytic activity of $WS_2$ nanosheets/nanodots ($WS_2$ NSDs) and activated $WS_2$ NSDs *versus* the $WS_2$-based catalysts on GCE (two catalysts on carbon fiber paper and carbon cloth have been pointed out) reported recently for HER in 0.5 M $H_2SO_4$ .

| Catalysts | Mass loading （mg cm$^{-2}$） | Overpotential (mV) for j=10 mA cm$^{-2}$ | Tafel Slope (mV dec$^{-1}$) | Reference |
|---|---|---|---|---|
| $WS_2$ NSDs | 0.283 | 337 | 80 | This work |
| Activated $WS_2$ NSDs | 0.283 | 255 | 73 | This work |
| $WS_2$ nanoflakes | 1 | ~358 | ~200 | 1 |
| BuLi exfoliated $WS_2$ nanosheets (~80% 1T- $WS_2$) | 0.001-0.0065 | 240 (1T) 440 (2H) | 55(1T) 110(2H) | 2 |
| BuLi exfoliated $WS_2$ | 0.0707 | ~690 | ~110 | 3 |
| $WS_{2(1-x)}Se_{2x}$ nanotubes on CFP | 0.21 | ~270 | 105 | 4 |
| $WS_2$ on carbon cloth | - | 225 | 105 | 5 |
| $WS_2$ nanosheets | 0.0566 | ~380 | ~95 | 6 |
| $WS_2$ nanosheets/quantum dots | 0.0354 | ~340 (DMF) | 70 (DMF) | 7 |

| | | ~355 (NMP) | 75 (NMP) | |
|---|---|---|---|---|
| Aromatic-exfoliated $WS_2$ | 0.0142 | ~520 | ~70 | 8 |
| $WS_{3-x}$ Films | - | 494 | 43.7 | 9 |
| Ta-doped $WS_2$ | 0.0707 | ~720 | ~170 | 10 |

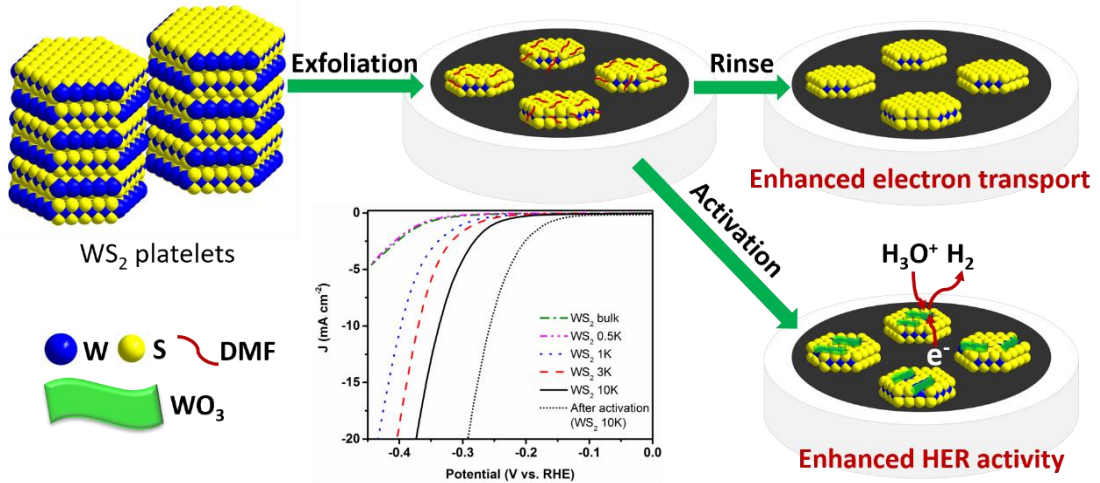